\documentclass[10pt,a4paper,twoside]{article}

\usepackage{fancyhdr}            
\usepackage{multicol}            
\usepackage{upgreek}             
\usepackage{indentfirst}         
\usepackage{booktabs}            
\usepackage{array,tabularx}      
\usepackage{keyval,graphicx}     
\usepackage{textcomp}            
\usepackage{amssymb,bm,mathrsfs} 
\usepackage[tbtags]{amsmath}     
\usepackage{lastpage}            
\usepackage[numbers,sort&compress]{natbib}  

\newcommand{\xiaosihao}{\fontsize{12.25pt}{\baselineskip}\selectfont}

\newcommand{\xiaowuhao}{\fontsize{8.5pt}{.5\baselineskip}\selectfont}
\newcommand{\liuhao}{\fontsize{8pt}{.6\baselineskip}\selectfont}

\newcommand{\newsmall}{\fontsize{9pt}{0.6\baselineskip}\selectfont}

\DeclareMathSizes{10}{10}{6}{5}   

\newcommand{\emark}[1]{$^{#1}$}

\newcommand{\thanksmark}{\textsuperscript{\,\rm{*}}}                          

\renewcommand{\,}{\hspace{0.125em plus 0.025em minus 0.025em}}

\newcommand{\oa}{\hbox{\fontfamily{ptm}\selectfont @}}                          

\newcommand{\ruledown}{\hfill\noindent{\lower.38cm\hbox{\rule{0.2pt}{0.4cm}}\rule{8.35cm}     
              {0.6pt}}\vspace*{-0.5cm}}

\let\asas=\cite
\renewcommand\cite[1]{\raisebox{0.5mm}{\textsuperscript{\asas{#1}}}}    

\makeatletter

\renewcommand{\thefootnote}{\fnsymbol{footnote}}    

\renewcommand{\thanks}[1]{\thanksmark
    \protected@xdef\@thanks{\@thanks
        \protect\footnotetext[0]{\hspace*{-6pt}*\,#1}}}

\newcounter{email}                        
\newcommand{\email}[1]{%
    \protected@xdef\@thanks{\@thanks%
        \protect\footnotetext[0]{\hspace*{-8pt}\arabic{email})\,{E-mail:\,}#1}}%
        \stepcounter{email}}%

\renewcommand\footnoterule{
  \kern 1\p@
  \hrule \@width37mm
  \kern 8\p@}

\renewcommand\@makefntext[1]{%
    \parindent 1em%
    \noindent
    \hb@xt@2em{\hss\@makefnmark}#1}

\renewcommand\maketitle{\par
  \begingroup
    \renewcommand\thefootnote{\@fnsymbol\c@footnote}%
    \def\@makefnmark{\rlap{\@textsuperscript{\normalfont\@thefnmark}}}%
    \long\def\@makefntext##1{\parindent 1em\noindent
            \hb@xt@2em{%
                \hss\@textsuperscript{\normalfont\@thefnmark}}##1}%
    \if@twocolumn
      \ifnum \col@number=\@ne
        \@maketitle
      \else
        \twocolumn[\@maketitle]%
      \fi
    \else
      \newpage
      \global\@topnum\z@   
      \@maketitle
    \fi
  \@thanks
  \endgroup
  \setcounter{footnote}{0}%
  \global\let\thanks\relax
  \global\let\maketitle\relax
  \global\let\@maketitle\relax
  \global\let\@thanks\@empty
  \global\let\@author\@empty
  \global\let\@date\@empty
  \global\let\@title\@empty
  \global\let\title\relax
  \global\let\author\relax
  \global\let\date\relax
  \global\let\and\relax}

\renewcommand\@maketitle{%
  \begin{center}%
  \let \footnote \thanks
   \vspace*{0.5em}
    {\Large\bf \@title \par}%
    {\normalsize
      \lineskip .5em%
      \vskip1.5em%
      \begin{tabular}[t]{c}%
        \@author%
      \end{tabular}}%
  \end{center}}%
\makeatother

\newcommand{\danwei}[1]{%
  \begin{center}%
    \vskip -1.25em%
    \liuhao%
    \begin{tabular}[t]{c}%
      #1%
    \end{tabular}%
  \end{center}%
}%

\renewenvironment{abstract}%
  {\small\vspace{0.5mm}%
   \list{}{\rightmargin\leftmargin}%
    \item{}{\bf Abstract}\hspace*{0.5em}\relax}%
   {\endlist}

\newenvironment{keyword}%
  {\small\vspace{1mm}%
    \list{}{\rightmargin\leftmargin}%
                \item{}{\bf Key~words}\hspace*{0.5em}\relax }%
       {\endlist%
        \vskip 8mm}%

\newcommand{\acknowledgments}[1]{%
  \par\vskip 8pt {\it #1}%
  }
\makeatletter

\renewcommand \thesection {\bf\@arabic\c@section}  

\renewcommand\section{\@startsection {section}{1}{\z@}%
                                    {5mm \@plus.2ex \@minus .2ex}%
                                   {5mm \@plus.2ex \@minus .2ex}%
                                   {\normalfont\large\bfseries}}
\renewcommand\subsection{\@startsection{subsection}{2}{\z@}%
                                     {1.5ex \@plus .2ex}%
                                     {1.5ex \@plus .2ex}%
                                     {\normalfont\bfseries}}

\renewcommand{\@biblabel}[1]{#1}
\renewcommand\refname{{\normalsize\bf References}}
\renewenvironment{thebibliography}[1]
     {\noindent\refname%
      \@mkboth{\MakeUppercase\refname}{\MakeUppercase\refname}%
      \liuhao
      \list{\@biblabel{\@arabic\c@enumiv}}%
           {\settowidth\labelwidth{\@biblabel{#1}}%
            \leftmargin\labelwidth
            \advance\leftmargin\labelsep
            \@openbib@code
            \usecounter{enumiv}%
            \let\p@enumiv\@empty
            \renewcommand\theenumiv{\@arabic\c@enumiv}}%
      \setlength{\itemsep}{-1.5mm}
      \setlength{\labelsep}{0.8em}
      \clubpenalty4000
      \@clubpenalty \clubpenalty
      \widowpenalty4000%
      \sfcode`\.\@m}
     {\def\@noitemerr
       {\@latex@warning{Empty `thebibliography' environment}}%
      \endlist}

\newenvironment{mylabc}
                {%
                 \newsmall
                 \let\\\@centercr
                 \list{}{\itemsep      \z@
                         \itemindent   -1em%
                         \listparindent0em
                         \leftmargin   3em
                         \rightmargin  2em}
                         \item\relax}
                {\endlist}
\makeatother

\newcommand{\ctitle}[1]{%
   \begin{center}
     \let \thanks \footnote%
     \vspace*{3mm}%
     {\LARGE\hei #1}%
     \vspace{2mm}%
   \end{center}}

\newcommand{\cauthor}[1]{%
   \begin{center}
     \setcounter{email}{1}
     \let \thanks \footnote%
     {\xiaosihao \fs #1}%
     \protect\par
     \vspace{4mm}
   \end{center}}

\fancypagestyle{myfoot}
{%
\fancyhf{}
\fancyhead[l]{\footnotesize\hspace{1em} \;xx\; \;  \;xx\;  \\[2pt] \hspace{1em}200x\; \;xx\;  }
\fancyhead[r]{\footnotesize Vol.\;xx, No.\;xx\mbox{\hspace{1em}}\\[2pt] xxx.,\;200x\hspace{1em}}
\fancyhead[c]{\hspace{0.5em} \hspace{0.5em} \hspace{0.5em} 
\hspace{0.5em} \hspace{0.5em} \hspace{0.5em} \hspace{0.5em} \\[2pt]
    \footnotesize HIGH\quad ENERGY\quad PHYSICS\quad AND\quad NUCLEAR\quad PHYSICS}
\fancyfoot[C]{}

}%
\makeatletter
\newcommand\capt[1]{%
\sbox\@tempboxa{\newsmall #1}%
\ifdim \wd\@tempboxa >\hsize
\begin{mylabc}
\vspace{-2mm}
{\newsmall #1}%
\vskip 1mm%
\end{mylabc}
\else
\global \@minipagefalse
\vspace*{-2mm}
\hb@xt@\hsize{\hfil\box\@tempboxa\hfil}%
\vskip 1mm%
\fi}
\makeatother
  
\usepackage{epsf,epsfig,latexsym}                
\begin{document}

\setcounter{page}{1}

\setlength{\abovecaptionskip}{4pt plus1pt minus1pt}
\setlength{\belowcaptionskip}{4pt plus1pt minus1pt}
\setlength{\abovedisplayskip}{6pt plus1pt minus1pt}
\setlength{\belowdisplayskip}{6pt plus1pt minus1pt}
\addtolength{\thinmuskip}{-1mu}
\addtolength{\medmuskip}{-2mu}
\addtolength{\thickmuskip}{-2mu}
\setlength{\belowrulesep}{0pt}
\setlength{\aboverulesep}{0pt}
\fancyhead[co]{\xiaowuhao }
\footnotetext[0]{Received 31 January 2007}

\title{Differential elliptic flow  at forward 
rapidity in 200 GeV AuAu collisions
\thanks{This work was supported in part
by the Office of Nuclear Physics of the U.S. Department of Energy under
contract DE-FG03-96ER40981.}}

\author{
    S.J. Sanders\emark{1)} (for the BRAHMS Collaboration)\email{ssanders\oa ku.edu}
}

\maketitle

\danwei{
1~(The University of Kansas, Lawrence, Kansas 66045  USA)\\
}

\begin{abstract}

Identified particle elliptic flow results are presented for the Au+Au reaction at
$\sqrt{s_{NN}} = 200$~GeV as a function
of transverse momentum and pseudorapidity. Data at pseudorapidities
$\eta \approx$ 0, 1, and 3.4 were obtained using the two BRAHMS spectrometers.   
Differential  $v_{2} (\eta ,p_{t})$ values for a given particle type are 
found to be essentially constant over the
covered pseudorapidity range, in contrast to the integral $v_{2}$ values
which have previously been observed to decrease at forward rapidities.
A softening of the particle spectra at forward angles is found to account for
at least part of the integral $v_{2}$ falloff. The data are found to be consistent
with existing constituent quark scaling systematics.  

\end{abstract}

\begin{keyword}
elliptic flow, rapidity dependence
\end{keyword}

\footnotetext[0]{\hspace*{-2em}\small\centerline{\thepage\ --- \pageref{LastPage}}}

\begin{multicols}{2}

\section{Introduction}

The azimuthal anisotropy of particle production in relativistic heavy-ion reactions is
believed to reflect pressure gradients of the thermalized medium formed early
in the collisions. 
In a Fourier expansion of the azimuthal behavior, the $2^\mathrm{nd}$-harmonic term, with
coefficient $v_{2}$, defines the ``elliptic flow'' of the reaction.    
Measurements near mid-rapidity at RHIC energies have shown that 
elliptic flow for particles below 2 GeV is close to that
expected from hydrodynamic models that assume the formation of a strongly
interacting quark-gluon plasma\cite{Shuryak:2003xe,Adams:2004bi,Adler:2003kt}. 
However, integral $v_{2}$ values, which are based solely on the asymmetry of the
overall  charged-particle production, are  found to  fall 
off going away from mid-rapidity, dropping by
about 35\% by pseudorapidity $\eta = 3$\cite{Back:2004zg}.
   
BRAHMS has now measured $v_{2}$ and the associated particle spectra 
as a function of transverse momentum 
for pions, kaons, and protons at angles (pseudorapidities)
$90^\mathrm{o} (\eta = 0)$, $40^\mathrm{o} (\eta = 1)$,
and $4^\mathrm{o} (\eta = 3.4)$.  The analysis of the elliptic flow behavior 
together with the particle spectra makes it possible to determine the extent 
to which the fall off of the integral $v_{2}$ values can be attributed to a
softening of the particle spectra going to forward rapidity.    

\section{Experimental Details}

Elliptic flow measurements characterize the azimuthal dependence of the
particle production with respect to the reaction plane. For any given collision,
an empirical reaction plane can be established by measuring the particle yield
observed in detectors mounted symmetrically about the beam axis.
BRAHMS used several different detector systems for this purpose:
Two rings of modestly segmented Si strip detectors (7 segments per strip)
were configured with 42 active segments surrounding the
beam axis in each ring.   Another ring using the same type of Si strip detector was configured
with six segments around the axis, and seven segments along the axis.  A scintillator tile 
ring with six panels was also arranged around the beam axis. Depending on the location
of the beam vertex, for collisions within $\pm 20$cm of the nominal vertex location  the Si and tile
detectors covered pseudorapidity values from -3 to +1.5.  At a 
somewhat larger rapidity ($\approx -3$), Cherenkov
radiators mounted to phototubes in one of the experiment's Beam-Beam counter 
arrays\cite{Adamczyk:2003sq} were also used for the reaction-plane determination. 
The general procedure for determining
the reaction plane and the corresponding reaction-plane correction factor
is described by Poskanzer and Voloshin\cite{Poskanzer:1998yz}. The
reaction plane corresponding to the second moment of the angular
distribution is found with
\begin{equation}
\Psi _2  = {1 \over 2}\sum\limits_i {{{w_i \sin \left( {2\phi _i } \right)} 
\over {w_i \cos \left( {2\phi _i } \right)}}} ,
\end{equation}
where the sum is over all detector elements with geometric weighting
factors $w_{i}$ and azimuthal angle $\phi _{i}$. 

\end{multicols}
\begin{figure}[htb]	
\centerline{\hbox{
\epsfxsize=12.0cm
\epsffile{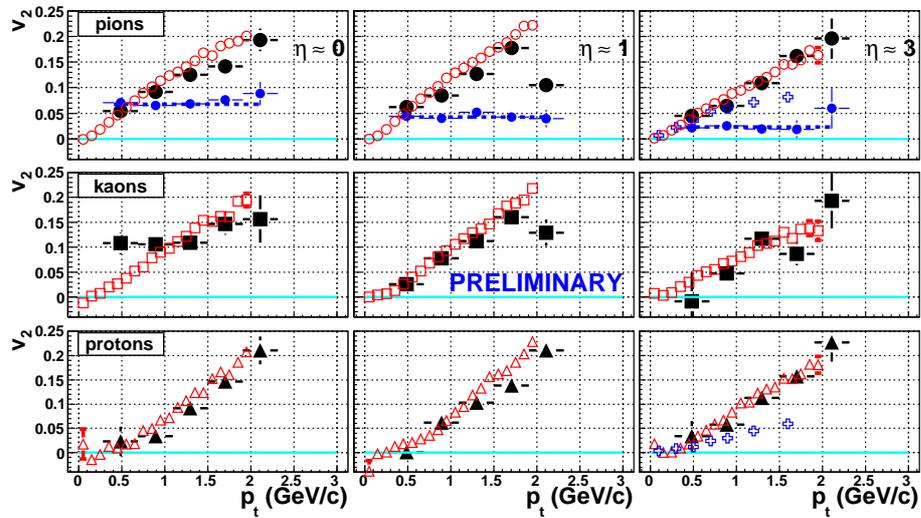}
}
}
\caption{
$\mathrm{v_{2}(p_{t}, \eta )}$ for pions (both charges), kaons (both charges), 
and protons (filled symbols).
Theoretical values based on the hydrodynamic calculations of Hirano 
{\it et al.}~[9] 
are shown by the open circles, squares, and triangles.  The
open crosses correspond to  $\mathrm{\eta \approx 4}$ calculations using
the AMPT model~[10]. The smaller (blue in color) 
points in the pion panels
are from a mixed-event analysis, as discussed in the text.} 
\label{fig:fig1}
\end{figure}
\begin{multicols}{2}

The angular correlation with respect to the
reaction plane of particles detected in the 
BRAHMS mid-rapidity and forward spectrometers  determine the observed $v_{2}$
values, with
\begin{equation}
v_2^{obs}  = \left\langle {\cos \left( {2\left[ {\phi  - \Psi _2 } \right]} \right)} \right\rangle .
\end{equation} 
Here the angular bracket denotes an average over all events of a given class,
such as all pions within a given range of transverse momenta. 
Since the BRAHMS spectrometers are small acceptance devices,
all particles detected in the mid-rapidity spectrometer have 
$\phi \approx 0^{o}$, while those detected by the forward spectrometer have 
$\phi \approx 180^{o}$.
 
The true $v_{2}$ value, which is based on the actual reaction plane,
is found from the $v_{2}^{obs}$ value using the reaction-plane correction
factor R, with
\begin{equation}
v_2  = v_2^{obs} /R  .
\end{equation}
The reaction-plane correction for any given
ring (or ring combination) is determined using two additional 
rings that
are far enough apart so as to avoid autocorrelations.  Considering
rings A, B, and C, the reaction-plane correction factor for ring A
is found as:
\begin{equation}
{\rm{R}}_{\rm{A}} {\rm{ = }}\sqrt {{{\left\langle {\cos \left( 
{2\left[ {\Psi _2^A  - \Psi _2^B } \right]} \right)} \right\rangle 
\left\langle {\cos \left( {2\left[ {\Psi _2^A  - \Psi _2^C } \right]} 
\right)} \right\rangle } \over {\left\langle {\cos \left( {2\left[ 
{\Psi _2^B  - \Psi _2^C } \right]} \right)} \right\rangle }}}. 
\end{equation}
The reaction-plane correction factor was typically in the range
$0.18 < R < 0.25$.

Figure~\ref{fig:fig1} shows the resulting values (closed symbols) 
of $v_{2}(p_{t})$ for pions, kaons, and protons at the indicated pseudorapidities, 
selecting events in the 10\% - 50\% centrality class. 
The behavior for the three different particle species shows very little
change with pseudorapidity.  The smaller (blue in color) filled circles shown in the pion
panels are the results of a mixed event analysis, where the reaction-plane angle used for the
analysis of a given identified pion is taken as that obtained from the previous event where a
particle of any type is detected in the same spectrometer.  Because of the limited angular acceptance
of the spectrometers, the deduced $v_{2}(p_{t})$ values from the mixed-event analysis
reflect the integral $v_{2}$ behavior.  The
scatter of the points with $p_{t}$ mirrors the statistical uncertainty in the measurements.  The fall
off of the average behavior, as indicated by the dashed lines, is consistent with the known fall-off of the
integral $v_{2}$ values going to forward rapidity.

\begin{figure}[htbp]
\centerline{\hbox{
\epsfxsize=6.0cm
\epsffile{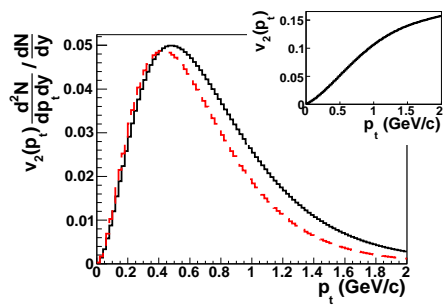}
}
}	
\caption{
Illustration of how the softening of the particle spectra going to forward
angles affects the integral $\mathrm{v_{2}}$ values.  The insert shows the
general behavior of the differential $\mathrm{v_{2}(p_{t})}$ values for pions,
which for this illustration is assumed to be independent of 
pseudorapidity.  Folding this distribution with the corresponding 
experimentally observed pion
spectra at $4^{o}$ and $90^{o}$ results in the dashed and solid curves, 
with integral $\mathrm{v_{2}}$ values of 0.046 and 0.036, respectively.
}
\label{fig:fig2}
\end{figure}

We have also
measured the $p_{t}$-dependent particle spectra for the channels where the
differential elliptic flow has been determined.
The average value of transverse momentum for pions in the
10\% - 50\% centrality class at $90^\mathrm{o}$ is
($475\pm 60$)~MeV/c, falling to ($380\pm 45$)~MeV/c at $4^\mathrm{o}$. 
These values are obtained by fitting a power-law dependence to the respective
particle spectra.
The softening of the particle spectrum going to more forward rapidities 
can have a significant effect on the integral $v_{2}$ values, as illustrated
in Figure~\ref{fig:fig2}. The insert shows an assumed form of $v_{2}(p_{t})$ 
that is
taken as being constant as a function of pseudorapidity.  
Folding this behavior with
the experimentally observed normalized particle spectra for pions at 
$90^\mathrm{o}$ 
and $4^\mathrm{o}$ results in the
solid and dashed lines of the main figure panel, respectively. The softer spectrum 
at the more forward
pseudorapidity leads to the weighted $v_{2}$ distribution peaking at a lower
mean $p_{t}$ value, resulting in a 22\% smaller integral $v_{2}$ 
value than a mid-rapidity. 
Two-thirds of the observed integral $v_{2}$ change is then 
attributable to the softening of the particle spectrum.     

\end{multicols}

\begin{figure}[htbp]
\centerline{\hbox{
\epsfxsize=12.0cm
\epsffile{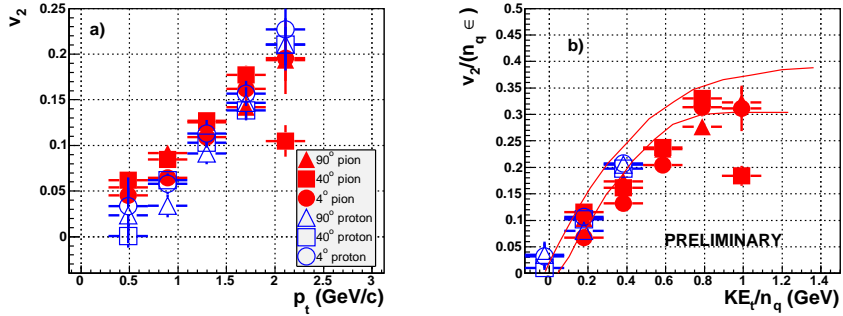}
}
}	
\caption{
a) $v_{2}(p_{t})$ vs. $p_{t}$ for pions and protons at $\eta$=0,1, and 3.
b) Constituent quark scaled values, as discussed in the text.  The lines
indicates the band containing most of the previous experimental results
near mid-rapidity.
}
\label{fig:fig3}
\end{figure}

\begin{multicols}{2}
The small change in the differential $v_{2}$ signal going from mid- to forward
rapidity  suggests a longitudinally extended region for the medium 
produced in the collision.  Rapidity dependent changes  in the 
radial flow or other rescattering behavior of the
hadronic stage  might then account for most of the fall off of the
integral $v_{2}$ signal going to forward rapidities.  

Hydrodynamic calculations are able to reproduce the observed behavior.  
This is seen by the open circles in Figure~\ref{fig:fig1} which show the
results of the hybrid hydrodynamic calculations of Hirano {\it et al.} 
(\cite{Hirano:2005xf}, and private communication) that include
dissipative effects of the late hadronic expansion stage. 
Good
agreement is found with the experimental values.
The open crosses in Figure~\ref{fig:fig1} show the results of the AMPT
model for pseudorapidity $\eta = 4$. At this angle the string melting
mechanism that allows for a good reproduction of mid-rapidity results
has been turned off~(\cite{Chen:2004vh}, and private communication).   
Although the comparison is subject to poor matching of pseudorapidity,
there is a suggestion that the longitudinal extent of the produced
medium may be greater than that assumed in current AMPT calculations.

The mid-rapidity elliptic flow behavior for many different systems have been
found to follow similar constituent scaling behavior when the $v_{2}$ value per valence quark,
scaled by the inverse of the Glauber-model eccentricity, 
is plotted as a function of the mean transverse kinetic energy per valence quark~\cite{Lacey:2006pn}.  
It Figure~\ref{fig:fig3}a we show the $v_{2}$ {\it vs.} $p_{t}$ behavior of pions and protons
for the three measured angles on the same plot.  In Figure~\ref{fig:fig3}b
these same data are now shown in terms of their constituent-quark scaling behavior.  
The two solid curves mark the approximate limits established by the
previous measurements.   The current data show a very similar trend, although 
with a tendency to fall somewhat below the previous results.   This may reflect a
systematic error in the measurement of the reaction-plane correction factor or
in the assumed Glauber eccentricity which is calculated for the trigger-weighted 
center of the centrality range.  An error in either of these quantities will tend
to have a similar scaling effect on all of our results.

\section{Conclusions}
BRAHMS has measured $v_{2}(p_{t},\eta )$ for pions, kaons,
and protons at $\sqrt{s_{NN}}$ = 200~GeV for the Au+Au reaction.  The results 
indicate a longitudinally extended region is produced where
the eccentricity of the created medium and the corresponding pressure
gradients remain remarkably constant.  Hydrodynamic calculations 
with final stage dissipation are found to be in excellent 
agreement with the measured differential $v_{2}$ values. The data are consistent with
constituent quark scaling over the covered pseudorapidity range.  

\acknowledgments{}

\end{multicols}

\vspace{-2mm}
\centerline{\rule{80mm}{0.1pt}}
\vspace{2mm}

\begin{multicols}{2}

\end{multicols}

\clearpage
\end{document}